\documentclass[prb,twocolumn]{revtex4}%

\usepackage{graphicx}
\usepackage{amsmath}
\usepackage{ulem}
\usepackage{color}

\newcommand{\be}{\begin{equation}}
\newcommand{\ee}{\end{equation}}
\newcommand{\bea}{\begin{eqnarray*}}
\newcommand{\eea}{\end{eqnarray*}}
\newcommand{\bean}{\begin{eqnarray}}
\newcommand{\eean}{\end{eqnarray}}

\RequirePackage{color}\definecolor{RED}{rgb}{1,0,0}\definecolor{BLUE}{rgb}{0,0,1} 
\DeclareOldFontCommand{\sf}{\normalfont\sffamily}{\mathsf} 
\providecommand{\DIFaddbegin}{} 
\providecommand{\DIFaddend}{} 
\providecommand{\DIFdelbegin}{} 
\providecommand{\DIFdelend}{} 
\providecommand{\DIFaddbeginFL}{} 
\providecommand{\DIFaddendFL}{} 
\providecommand{\DIFdelbeginFL}{} 
\providecommand{\DIFdelendFL}{} 
\newcommand{\DIFscaledelfig}{0.5}
\RequirePackage{settobox} 
\RequirePackage{letltxmacro} 
\newsavebox{\DIFdelgraphicsbox} 
\newlength{\DIFdelgraphicswidth} 
\newlength{\DIFdelgraphicsheight} 
\LetLtxMacro{\DIFOincludegraphics}{\includegraphics} 
\newcommand{\DIFaddincludegraphics}[2][]{{\color{blue}\fbox{\DIFOincludegraphics[#1]{#2}}}} 
\newcommand{\DIFdelincludegraphics}[2][]{
\sbox{\DIFdelgraphicsbox}{\DIFOincludegraphics[#1]{#2}}
\settoboxwidth{\DIFdelgraphicswidth}{\DIFdelgraphicsbox} 
\settoboxtotalheight{\DIFdelgraphicsheight}{\DIFdelgraphicsbox} 
\scalebox{\DIFscaledelfig}{
\parbox[b]{\DIFdelgraphicswidth}{\usebox{\DIFdelgraphicsbox}\\[-\baselineskip] \rule{\DIFdelgraphicswidth}{0em}}\llap{\resizebox{\DIFdelgraphicswidth}{\DIFdelgraphicsheight}{
\setlength{\unitlength}{\DIFdelgraphicswidth}
\begin{picture}(1,1)
\thicklines\linethickness{2pt} 
{\color[rgb]{1,0,0}\put(0,0){\framebox(1,1){}}}
{\color[rgb]{1,0,0}\put(0,0){\line( 1,1){1}}}
{\color[rgb]{1,0,0}\put(0,1){\line(1,-1){1}}}
\end{picture}
}\hspace*{3pt}}} 
} 
\LetLtxMacro{\DIFOaddbegin}{\DIFaddbegin} 
\LetLtxMacro{\DIFOaddend}{\DIFaddend} 
\LetLtxMacro{\DIFOdelbegin}{\DIFdelbegin} 
\LetLtxMacro{\DIFOdelend}{\DIFdelend} 
\DeclareRobustCommand{\DIFaddbegin}{\DIFOaddbegin \let\includegraphics\DIFaddincludegraphics} 
\DeclareRobustCommand{\DIFaddend}{\DIFOaddend \let\includegraphics\DIFOincludegraphics} 
\DeclareRobustCommand{\DIFdelbegin}{\DIFOdelbegin \let\includegraphics\DIFdelincludegraphics} 
\DeclareRobustCommand{\DIFdelend}{\DIFOaddend \let\includegraphics\DIFOincludegraphics} 
\LetLtxMacro{\DIFOaddbeginFL}{\DIFaddbeginFL} 
\LetLtxMacro{\DIFOaddendFL}{\DIFaddendFL} 
\LetLtxMacro{\DIFOdelbeginFL}{\DIFdelbeginFL} 
\LetLtxMacro{\DIFOdelendFL}{\DIFdelendFL} 
\DeclareRobustCommand{\DIFaddbeginFL}{\DIFOaddbeginFL \let\includegraphics\DIFaddincludegraphics} 
\DeclareRobustCommand{\DIFaddendFL}{\DIFOaddendFL \let\includegraphics\DIFOincludegraphics} 
\DeclareRobustCommand{\DIFdelbeginFL}{\DIFOdelbeginFL \let\includegraphics\DIFdelincludegraphics} 
\DeclareRobustCommand{\DIFdelendFL}{\DIFOaddendFL \let\includegraphics\DIFOincludegraphics} 
\RequirePackage{listings} 
\RequirePackage{color} 
\lstdefinelanguage{DIFcode}{ 
  moredelim=[il][\color{red}\scriptsize]{\%DIF\ <\ }, 
  moredelim=[il][\color{blue}\sffamily]{\%DIF\ >\ } 
} 
\lstdefinestyle{DIFverbatimstyle}{ 
        language=DIFcode, 
        basicstyle=\ttfamily, 
        columns=fullflexible, 
        keepspaces=true 
} 
\lstnewenvironment{DIFverbatim}{\lstset{style=DIFverbatimstyle}}{} 
\lstnewenvironment{DIFverbatim*}{\lstset{style=DIFverbatimstyle,showspaces=true}}{} 

\begin{document}

\draft
\title{\bf Effects of metallic electrodes on the thermoelectric properties of zigzag graphene nanoribbons with periodic vacancies}

\author{David M T Kuo}

\address{Department of Electrical Engineering and Department of Physics, National Central
University, Chungli, 320 Taiwan}

\date{\today}

\begin{abstract}
{We theoretically analyze the thermoelectric properties of
graphene quantum dot arrays (GQDAs) with line- or
surface-contacted metal electrodes. Such GQDAs are realized as
zigzag graphene nanoribbons (ZGNRs) with periodic vacancies. Gaps
and minibands are formed in these GQDAs, which can have metallic
and semiconducting phases. The electronic states of  the first
conduction (valence) miniband with nonlinear dispersion may have
long coherent lengths along the zigzag edge direction. With
line-contacted metal electrodes, the GQDAs have the
characteristics of serially coupled quantum dots (SCQDs) if the
armchair edge atoms of the ZGNRs are coupled to the electrodes. By
contrast, the GQDAs have the characteristics of parallel QDs if
the zigzag edge atoms are coupled to the electrodes. The maximum
thermoelectric power factors of SCQDs with line-contacted
electrodes of Cu, Au, Pt, Pd, or Ti at room temperature were
similar or greater than $0.186nW/K$; their figures of merit were
greater than three. GQDAs with line-contacted metal electrodes
have much better thermoelectric performance than surface contacted
metal electrodes.}
\end{abstract}

\maketitle

\section{Introduction}
Charge transport through individual quantum dots with discrete
levels exhibits an interesting phenomenon such as the Kondo effect
and Coulomb blockade.[\onlinecite{HaugH}] Quantum dots (QDs) have
also been suggested as implementations of low-power devices. These
low-power devices include single-electron
transistors[\onlinecite{GuoLJ},\onlinecite{Postma}], single-photon
sources[\onlinecite{Michler}-\onlinecite{ChangWH}], single-photon
detectors[\onlinecite{Gustavsson}] and single-electron heat
engines[\onlinecite{Josefsson}]. Some applications of QD devices
require both high efficiency and considerable output power. These
applications require QD solids that can retain the size-tunable
properties of QDs and possess the band transport characteristic of
bulk semiconductors [\onlinecite{Kagan}]. Although substantial
effort has been devoted to producing such QD solids, experimental
studies of the thermoelectric properties of such one-dimensional
QD arrays are lacking [\onlinecite{Harman},\onlinecite{Talgorn}].
Dot-size fluctuation is a challenge for thermoelectric devices
made from 1D silicon QD arrays (QDAs) [\onlinecite{Lawrie}].

Recently, numerous studies have focused on two-dimensional (2D)
materials such as graphenes [\onlinecite{Novoselovks}], and 2D
transition metal dichalcogenides (TMDs) and oxides
[\onlinecite{Geim}--\onlinecite{Desai}]. Quasi-1D graphene
nanoribbons (GNRs) can now be fabricated with atomic precision by
using the bottom-up approach [\onlinecite{Cai}]; this approach can
yield complex graphene-based nanostructures on-demand for quantum
device applications [\onlinecite{LiuJ}--\onlinecite{Almeida}]. The
superlattices of zigzag GNRs (ZGNRs) behave as graphene QDAs
(GQDAs) [\onlinecite{Topsakal}]. The phonon thermal conductance of
GQDAs can be dramatically reduced by at least one order of
magnitude compared with that of ZGNRs [\onlinecite{Cuniberti}].
Therefore, GQDAs are expected to have promising applications in
nanoscale energy harvesting.

Nevertheless, previous studies have demonstrated that the contact
types of grapehene and GNRs significantly influence electron
transport between the electrodes
[\onlinecite{Darancet}--\onlinecite{Matsuda}]. To reduce the
number of electron backward scattering components, GNR devices can
be produced using normal metal electrodes instead of graphene
electrodes to improve the transmission coefficient
[\onlinecite{GLiang}]. In addition, the contact resistance of a
graphene with line-contacted metal electrodes can be much smaller
than that of a graphene with surface-contacted metal electrodes
[\onlinecite{Matsuda}]. Other 2D materials also have this feature
[\onlinecite{Shen}, \onlinecite{ChenRS}]. These results all
indicate that the interface properties arising from the presence
of a contact junction is a key factor affecting the implementation
of graphene-based or TMD-based electronic and thermoelectric
devices.

Few studies have examined how the performance of GQDA
thermoelectric devices is affected by the material properties and
geometries of various contacting metals. In this study, we
theoretically investigate the ballistic transport and
thermoelectric properties of GQDAs coupled to various metal
electrodes, as shown in Fig.~1; GQDAs are realized by ZGNRs with
periodic vacancies. We have focused on the channel length between
thermal contact shorter than the electron mean free path
($\lambda_e$) but longer than the phonon mean free path
($\lambda_{ph}$). We calculated the electrical conductance
($G_e$), Seebeck coefficient ($S$) and electron thermal
conductance ($\kappa_e$) for different line-contacted metal
electrodes and surface-contacted metal electrodes using the
tight-binding model and Green's function. We find that the maximum
thermoelectric power factor values of GQDAs with Cu, Au, Pt, Pd or
Ti line-contacted electrodes can reach up to $79\%$ of the power
factor of the 1D theoretical limit at room temperature.

\begin{figure}[h]
\centering
\includegraphics[trim=2.5cm 0cm 2.5cm 0cm,clip,angle=0,scale=0.3]{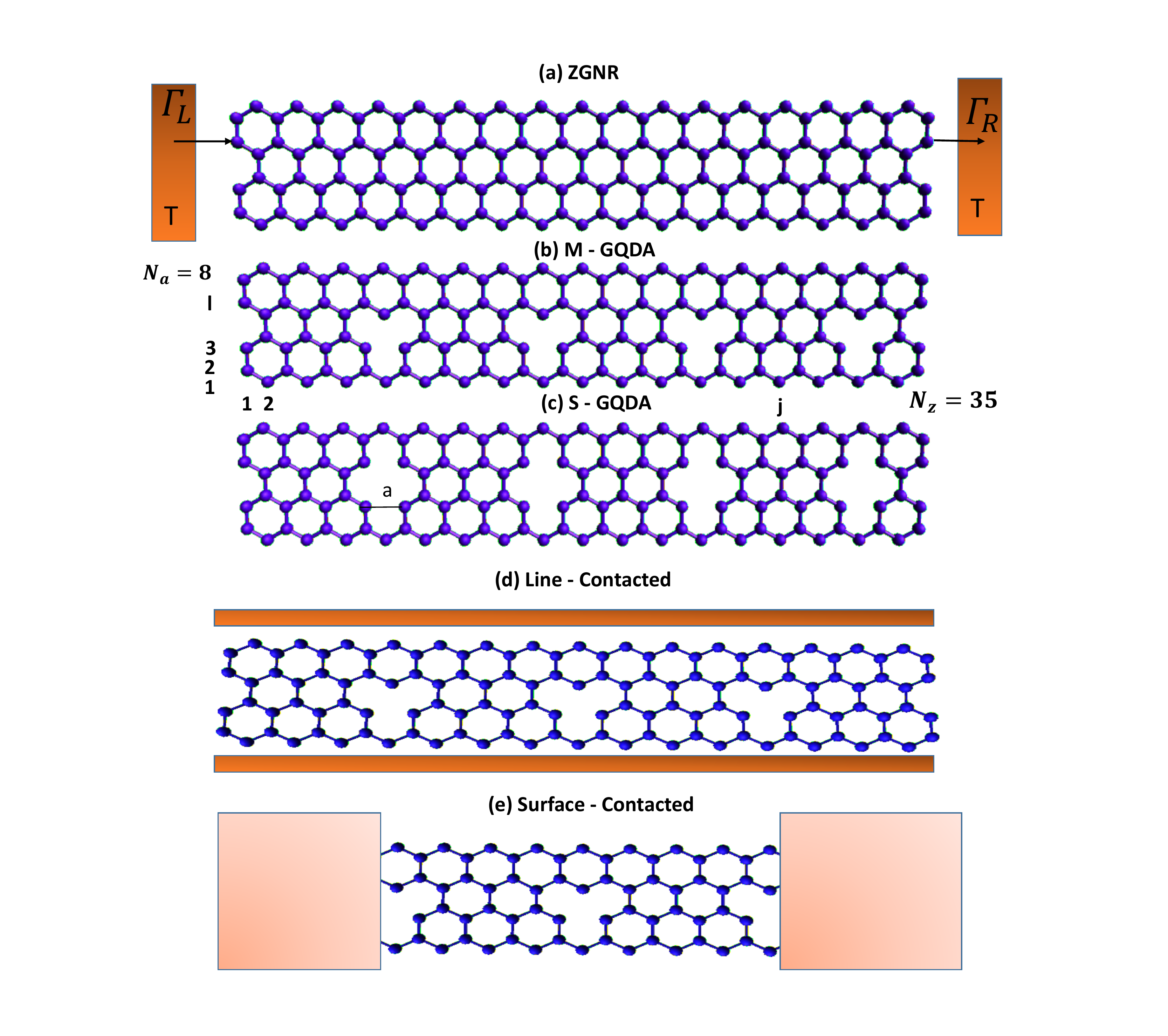}
\caption{Diagram of ZGNR without and with vacancies for $N_a=8$
and $N_z=35$. (a) Armchair edge atoms of ZGNRs coupled to metallic
electrodes. $\Gamma_{L}$ ($\Gamma_R$) denotes the tunneling rate
of the electrons between the left (right) electrode and the
leftmost (rightmost) carbon atoms of the armchair edges. $T$
denotes the equilibrium temperature of the left (right) electrode.
(b, c) Graphene quantum dot arrays (GQDAs) are realized by
periodically removing one carbon and two carbons from the interior
sites of ZGNRs. The distance between the nearest vacancies is
$D=4a$, where the lattice constant of graphene is $a=2.46 \AA$.
(d) The zigzag edge atoms of GQDAs are coupled to the metal
electrodes. (e) GQDAs with surface-contacted metal electrodes.}
\end{figure}

\section{Calculation method}
A good approximation of the electronic states of a GQDA realized
by ZGNRs with periodic vacancies is a tight-binding model with one
$p_z$ orbital per atomic site [\onlinecite{Topsakal},
\onlinecite{Wakabayashi2}]. The Hamiltonian of the GQDA, written,
$H_{GQDA}$, is

\begin{small}
\begin{eqnarray}
H_{GQDA}&= &\sum_{\ell,j} E_{\ell,j} d^{\dagger}_{\ell,j}d_{\ell,j}\\
\nonumber&-& \sum_{\ell,j}\sum_{\ell',j'} t_{(\ell,j),(\ell', j')}
d^{\dagger}_{\ell,j} d_{\ell',j'} + h.c,
\end{eqnarray}
\end{small}
where $E_{\ell,j}$ is the on-site energy for the $p_z$ orbital in
the ${\ell}$th row and $j$th column. $d^{\dagger}_{\ell,j}
(d_{\ell,j})$ creates (destroys) one electron at the atom site
labeled by ($\ell$,$j$) where $\ell$ and $j$ are the row and
column indices, respectively, as depicted in Fig.~1.
$t_{(\ell,j),(\ell', j')}$ describes the electron hopping energy
from site ($\ell$,$j$) to site ($\ell'$,$j'$). The tight-binding
parameters used for GQDAs are $E_{\ell,j}=0$ for the on-site
energy and $t_{(\ell,j),(\ell',j')}=t_{pp\pi}=2.7$ eV for the
nearest-neighbor hopping strength. Here, the electron--electron
Coulomb interactions are neglected. Their effect is discussed in a
subsequent section.

Thermoelectric coefficients, namely the electrical conductance
($G_e$), Seebeck coefficient ($S$) and electron thermal
conductance ($\kappa_e$) were calculated using $G_e=e^2{\cal
L}_{0}$, $S=-{\cal L}_{1}/(eT{\cal L}_{0})$ and
$\kappa_e=\frac{1}{T}({\cal L}_2-{\cal L}^2_1/{\cal L}_0)$ with
${\cal L}_n$ ($n=0,1,2$) defined as
\begin{equation}
{\cal L}_n=\frac{2}{h}\int d\varepsilon~ {\cal
T}_{LR}(\varepsilon)(\varepsilon-\mu)^n\frac{\partial
f(\varepsilon)}{\partial \mu}.
\end{equation}
Here, $f(\varepsilon)=1/(exp^{(\varepsilon-\mu)/k_BT}+1)$ is the
Fermi distribution function of electrodes at equilibrium
temperature $T$ and chemical potential $\mu$. ${\cal
T}_{LR}(\varepsilon)$ denotes the transmission coefficient of a
GQDA connected to electrodes, which can be calculated using the
formula ${\cal
T}_{LR}(\varepsilon)=4Tr[\Gamma_{L}(\varepsilon)G^{r}(\varepsilon)\Gamma_{R}(\varepsilon)G^{a}(\varepsilon)]$
[\onlinecite{KuoDM}--\onlinecite{KuoD}], where
($\Gamma_{L}(\varepsilon)$ and $\Gamma_{R}(\varepsilon)$) denote
the tunneling rate at the left and right leads, respectively, and
{${G}^{r}(\varepsilon)$ and ${G}^{a}(\varepsilon)$ are the
retarded and advanced Green's functions of the GQDA, respectively.
The tunneling rates are described by the imaginary part of the
self energy resulting from the coupling between the left (right)
electrode with its adjacent GQDA atoms. Based on the tight-binding
orbitals, $\Gamma_{\alpha}(\varepsilon)$ and Green's functions are
matrices. For simplicity, $\Gamma_{\alpha}$ for interface carbon
atoms has diagonal entries given by the same constant $\Gamma_t$.
The magnitudes of $\Gamma_t$ depend on the line- or
surface-contacted metal electrodes [\onlinecite{Matsuda}]. The
thermoelectric efficiency of graphene quantum dot arrays (GQDAs)
is determined by the dimensionless figure of merit
$ZT=S^2G_eT/(\kappa_e+\kappa_{ph})$, where $\kappa_{ph}$ denotes
the phonon thermal conductance of GQDAs. First-principle
calculations of $\kappa_{ph}$ are beyond the scope of this
article, so for simplicity, we calculate $\kappa_{ph}$ using an
empirical method described in ref. [\onlinecite{KuoD}]. The phonon
mean free path $\lambda_{ph}$ has been measured to be reduced from
$300-600$ nm in a single-layer graphene to $10$ nm in graphene
nanostructures (see [\onlinecite{XuY}] and references therein). In
this study, we chose zigzag graphene nanoribbons (ZGNRs) with $N_z
= 127$ ($L_z = 15.5$ nm) because the channel length between
thermal contacts satisfies the condition of being shorter than
$\lambda_e$ but longer than $\lambda_{ph}$.

\section{Results and discussion}
\subsection{Electronic energy levels}
Graphene with gapless characteristics has limited applications in
devices [\onlinecite{Nakada}, \onlinecite{Wakabayashi}]. However,
graphene with periodic vacancies has minibands, which enable its
applications in thermoelectric devices at low temperatures
[\onlinecite{Gunst},\onlinecite{ChangPH}]. To enable the
graphene-based devices operating at room temperature, some studies
have considered ZGNRs with periodic vacancies
[\onlinecite{ZhangYT}]. However, few studies have investigated how
line- and surface-contacted metal electrodes affect the
thermoelectric properties of ZGNRs with vacancies. The electron
band structures of infinite ZGNRs without and with periodic
vacancies are presented in Fig. 2. The band structures of infinite
ZGNRs with various widths ($N_a$) have been extensively studied
[\onlinecite{Nakada}]. The localized zero-energy flat-band modes
of the first subband are from $k = \frac{2\pi}{3a}$ to
$\frac{\pi}{a}$ as $N_a \rightarrow \infty$. In Fig. 2(a), the
states of the first subband are the localized states for $k$
within $k_c=0.738 \frac{\pi}{a} \le k \le \frac{\pi}{a}$, where
$k_c$ is determined by \DIFdelbegin
$k_c/2=\textrm{arccos}(N_a/(2N_a+4))$  and $N_a=8$
[\onlinecite{Wakabayashi2}]. The edge state with the shortest
decay length along the armchair direction occurs at
$k=\frac{\pi}{a}$. When $k$ deviates from $\frac{\pi}{a}$, the
zigzag edges states form bonding and antibonding states, and the
zero-energy modes disappear [Fig. 2(a)] . When $|E| \ge 2.16$ eV,
we observe the second conduction and valence subbands. Because
these are bulk states, this study focuses on the edge states of
the first subband with $|E| \le 2.16$ eV. For ZGNRs with periodic
vacancies [Fig. 2(b) and Fig. 2(c)], the gapped energy regions are
open and minibands are formed. ZGNR with periodic vacancies can be
in the semiconducting phase as illustrated by the sizable central
gap (0.959 eV) in Fig.~2(c). To reflect this phase difference,
they are denoted metal GQDAs (M-GQDAs) and semiconductor GQDAs
(S-GQDAs}). Because the  finite channel length is shorter than
$\lambda_e$ but longer than $\lambda_{ph}$, the length-dependent
energy levels of GQDAs should be further clarified.

\begin{figure}[h]
\centering
\includegraphics[angle=0,scale=0.3]{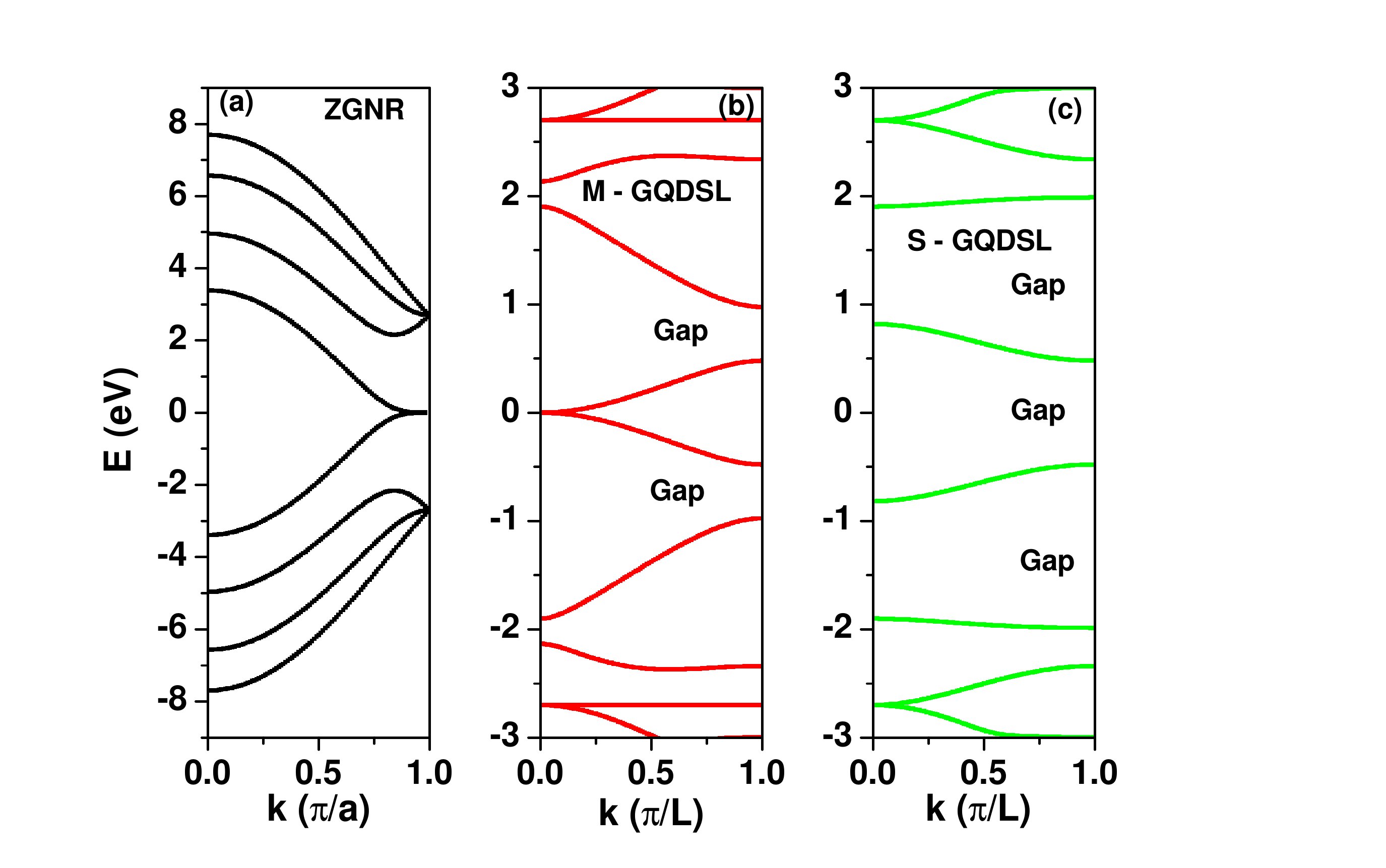}
\caption{Electronic band structures of infinite ZGNRs (a) without
and (b, c) with periodic vacancies. The distance between the
nearest vacancies is always $L=4a$ in this article. The size of a
super unit cell is characterized by $N_a=8$ and $N_z=8$.}
\end{figure}

To illustrate the characteristics of the energy levels in each
miniband of Fig. 2(b) and 2(c), we present the calculated
eigenvalues as functions of QD number $N_{QD}$ (or
$N_z=8N_{QD}-1$) in Fig. 3. The spectra reveal the transformation
from the molecule-like state to band-like state as $N_{QD}$
increases from 3 toward infinity (Fig. 2). In Fig. 3, the number
of energy levels in each miniband is given by $N_{QD}$,
demonstrating that finite-size ZGNRs with periodic vacancies
exhibit the characteristics of GQDAs. A single GQD can be
characterized by the size of $N_a=8$ and $N_z=7$, which results in
four energy levels of $\varepsilon_{h,2}=-1.9024$ eV,
$\varepsilon_{h,1}=-0.47905$ eV, $\varepsilon_{e,1}=0.47905$ eV
and $\varepsilon_{e,2}=1.9024$ eV in the range of $|E| \le 2.16$
eV. These four energy levels do not vary with $N_{QD}$. To
understand this  behavior, we calculate the eigenvalues of ZGNRs
without periodic vacancies as a function of $ N_z=8N_{QD}-1$ for
$N_a=8$ and obtain $N_z$-independent $\varepsilon_{e(h),1}$ and
$\varepsilon_{e(h),2}$. The wave functions of
$\varepsilon_{e(h),1}$ and $\varepsilon_{e(h),2}$ of the ZGNRs are
vanishingly small at $j=8m$ (so-called nodes), where $m$ is an
integer. This explains the presence of $N_{QD}$-independent
$\varepsilon_{e(h),1}$ and $\varepsilon_{e(h),2}$ in Fig. 3. The
sinusoidal wave of $\varepsilon_{e(h),1}$ ($\varepsilon_{e(h),2}$)
along the zigzag edges has a much  longer coherence length than
away from these edges. Although the periodic vacancies do not
affect the energy levels of $\varepsilon_{e(h),1}$ and
$\varepsilon_{e(h),2}$, they restrict and perturb some energy
levels of ZGNRs. Therefore, they are called antidots
[\onlinecite{Gunst}--\onlinecite{ZhangYT}].

\begin{figure}[h]
\centering
\includegraphics[angle=0,scale=0.3]{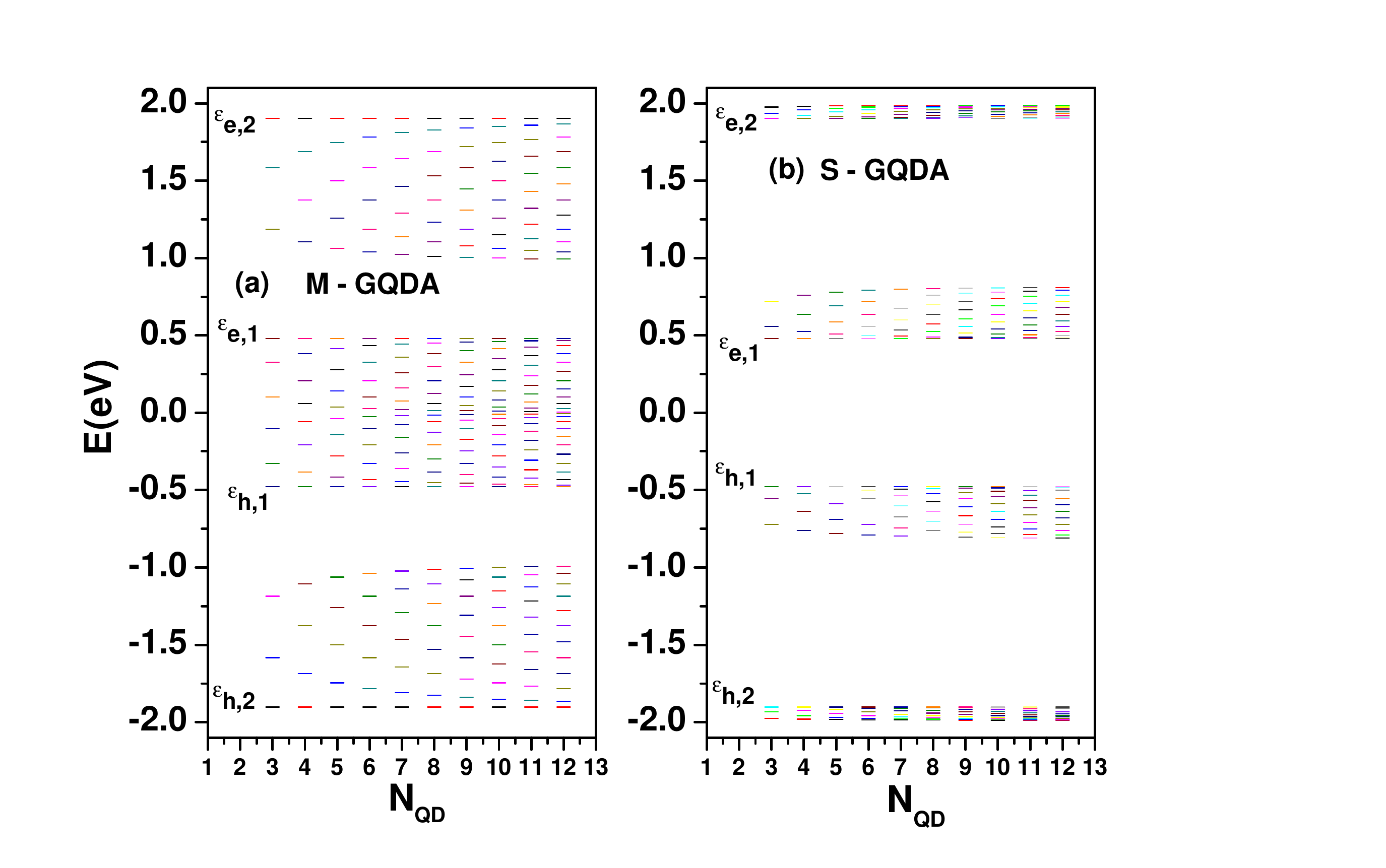}
\caption{Spectra of (a) M-GQDA and (b) S-GQDA as functions of
$N_{QD}$. Each GQD in the GQDA structures has size $N_a=8$ and
$N_z=7$. The length of the  GQDA is $N_z=8N_{QD}-1$.}
\end{figure}

\subsection{ Line-contacted metal electrodes}
When a graphene is coupled to metal electrodes, contact properties
such as the Schottky barrier or ohmic contact and contact
geometries can significantly affect the electron transport in
graphene [\onlinecite{Matsuda},
\onlinecite{ChuT}--\onlinecite{Hancheng}]. Although many
theoretical studies have attempted to clarify this key behavior
from first principles, only qualitative results regarding
$\Gamma_t$ arising from the contact junction can be obtained due
to theoretical limitations [\onlinecite{YangL},
\onlinecite{LeeG}]. In Fig. 4, we present that the calculated
transmission coefficient of interface $T_c(\varepsilon)$ based on
the configuration considered in [\onlinecite{Matsuda}], in which
the authors introduced ``super unit cells '' with eight unit cells
and investigated the armchair edge atoms of graphene coupled to
various  metal electrodes. Contact resistance $R_c=1/G_{e,c}$ per
unit cell is determined by $G_{e,c}=G_0T_c(\varepsilon=0)$, where
$G_0=2e^2/h=1/(12.9k\Omega)=77.5\mu S$ is the quantum conductance
and $T_c(\varepsilon=0)$ is the contact transmission coefficient
at the Fermi energy. In [\onlinecite{Matsuda}], the authors
reported that $R_c=13.3, 17.8, 18.6, 23.3$ and $31.7$ $k\Omega$
for Ti, Pd, Pt, Au and Cu, respectively. The $\Gamma_t$ of each
carbon atom at the interface can be obtained by fitting these
contact resistances. We obtain the $\Gamma_t$ values of Cu, Au,
Pt, Pd, and Ti as $\Gamma_t=0.495$ eV, $\Gamma_t=0.603$ eV,
$\Gamma_t=0.693$ eV, $\Gamma_t=0.72$ eV, and $\Gamma_t=0.9$ eV,
respectively.

\begin{figure}[h]
\centering
\includegraphics[angle=0,scale=0.3]{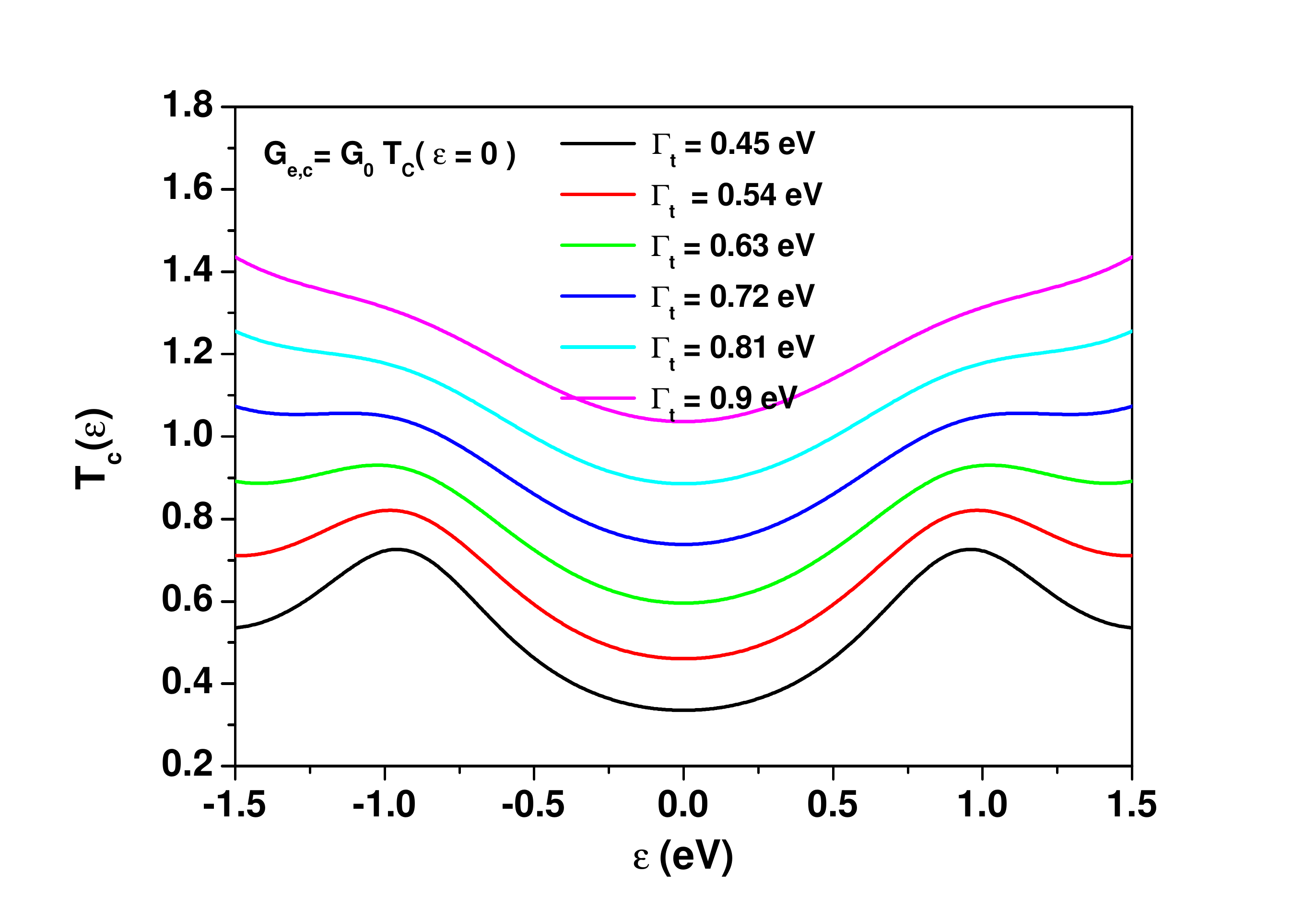}
\caption{Transmission coefficient at the metal--graphene interface
per unit cell for various $\Gamma_t$ values. Armchair edge atoms
of graphene are coupled to metal electrodes. A super unit cell
comprises 16 carbon atoms.}
\end{figure}

\subsubsection{Armchair edge atoms coupled to electrodes}
Based on the aforementioned $\Gamma_t$ values,  Fig.~5 presents
the calculated ${\cal T}_{LR}(\varepsilon)$ of M-GQDA with $N_a=8$
and $N_z=127$ for various metal electrodes. The maximum ${\cal
T}_{LR}(\varepsilon)= 1$ indicates that GQDAs behave as serially
coupled QDs (SCQDs). Note that two energy levels
$\varepsilon_{e(h)}=\pm 0.47905eV$ and $\varepsilon_{e(h)}=\pm
0.47179eV$ are too close to be resolved in the spectra of ${\cal
T}_{LR}(\varepsilon)$. From Eq. (2), the area below the ${\cal
T}_{LR}(\varepsilon)$ curve has a considerable effect on the
thermoelectric coefficients. As illustrated in Fig. 5, the area
below the ${\cal T}_{LR}(\varepsilon)$ curve increases with the
tunneling rate. Moreover, these areas maintain a near-right
triangle shape that exhibits a steep change with respect to
$\varepsilon$ on the sides of the $\varepsilon_{e,1}$ and
$\varepsilon_{h,1}$ edges. For energy harvesting applications at
room temperature, we must design a ${\cal T}_{LR}(\varepsilon)$
spectrum with a square form (SF) to obtain the optimal figure of
merit and electrical power output.[\onlinecite{Whitney}] However,
no method has been developed for realizing a ${\cal
T}_{LR}(\varepsilon)$ curve area with SF in a finite channel
length ($L_z < \lambda_e$) between thermal contacts
[\onlinecite{KuoD}].

\begin{figure}[h] \centering
\includegraphics[angle=0,scale=0.3]{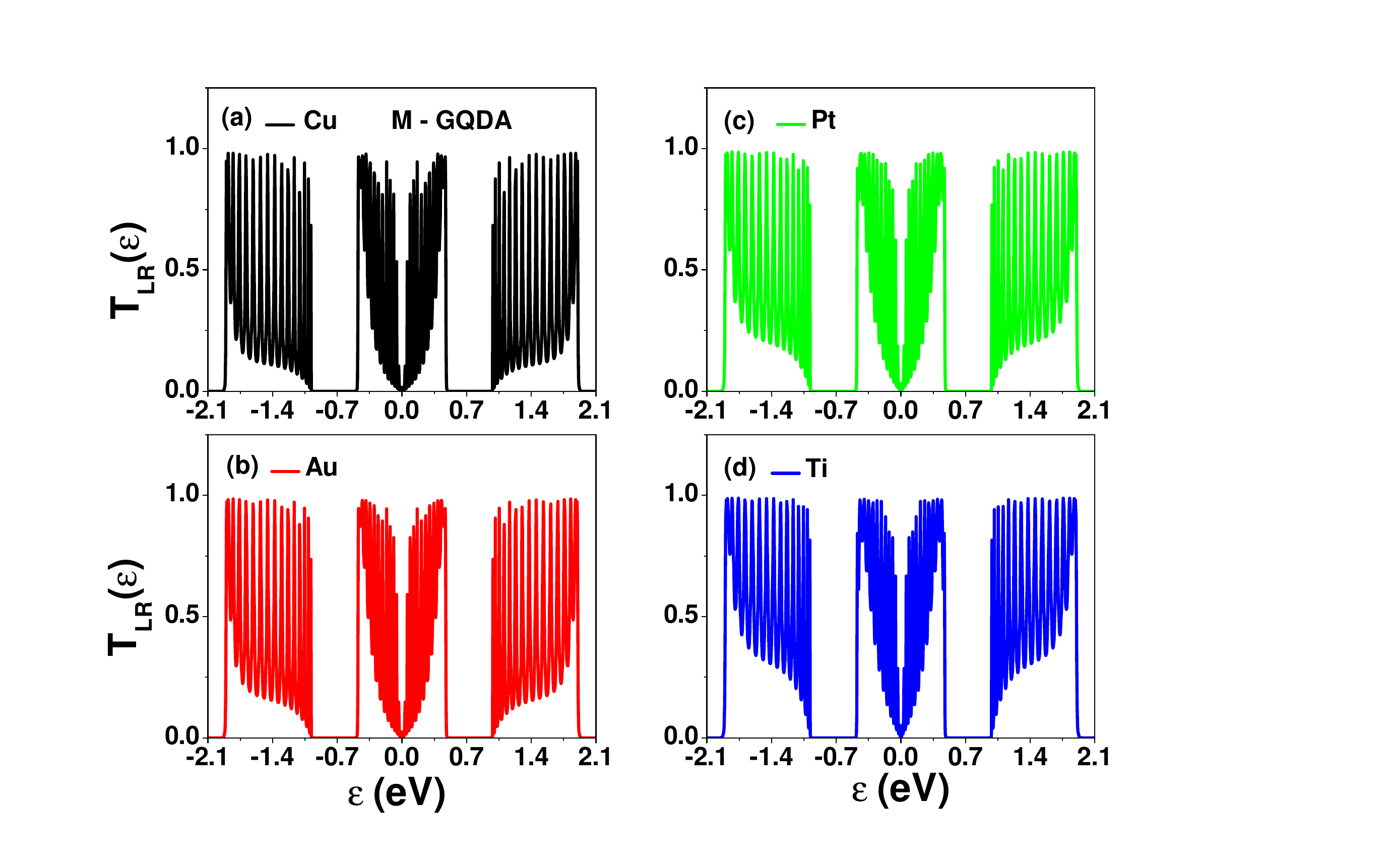}
\caption{Transmission coefficient ${\cal T}_{LR}(\varepsilon)$ of
a finite M-GQDA as functions of $\varepsilon$ for various
line-contacted metal electrodes at $N_a=8$ ( $L_a=0.71$nm ) and
$N_z=127$ ($L_z=15.5$nm ). Diagrams (a), (b), (c), and (d)
correspond to Cu, Au,Pt,and Ti, respectively. }
\end{figure}

We then calculate the electrical conductance ($G_e$), Seebeck
coefficient ($S$), power factor ($PF$), and figure of merit ($ZT$)
of M-GQDAs at room temperature (Fig. 6). $G_e$, $S$, and $PF$ are
in units of $G_0=2e^2/h$, $k_B/e=86.25\mu V/K$, and
$2k^2_B/h=0.575 pW/K^2$, respectively. Unlike $G_e$, $S$ is less
sensitive to variations in $\Gamma_t$ [Fig.6(b)]. The behavior of
$S$ can be approximated  by $S=-\frac{\pi^2k^2_BT}{3e}
\frac{\partial ln(G_e(\mu,T))}{\partial \mu}$. The maximum $PF$
and $ZT$ values are given by $\mu$ in the gap region. The trend of
$ZT$ with respect to various metal electrodes is determined by the
power factor ($PF=S^2G_e$) because  the thermal conductance of
GQDAs is dominated by phonon thermal conductance $\kappa_{ph}$. We
calculated the dimensionless $ZT=S^2G_eT/(\kappa_e+\kappa_{ph})$
with $\kappa_{ph}=F_s\kappa_{ph,ZGNR}$, where
$\kappa_{ph,ZGNR}=\frac{\pi^2k^2_BT}{3h}$ is the phonon thermal
conductance of an ideal ZGNR [\onlinecite{Zhengh}] and $F_s=0.1$
denotes a reduction factor resulting from periodic vacancies in
ZGNRs [\onlinecite{Cuniberti}, \onlinecite{ChangPH}]. The results
in Fig.~6 imply that $ZT > 3$ can be realized if GQDAs are
considered to be coupled to metallic electrodes. The four
investigated  metal electrodes have similar maximum $PF$ and $ZT$
values; the maximum $PF$ can reach $79\%$ of the theoretical limit
for 1D systems of $PF_{QB}=1.2659(\frac{2k^2_B}{h})$
[\onlinecite{Whitney}]. ZGNRs are graphene-based 1D topological
insulators (TIs), and $ZT$ of 1D TIs greater than  3 has been
theoretically demonstrated in the absence of periodic vacancies
[\onlinecite{Xu}]. The $ZT$ values of 1D TIs tend to be at their
maximum when  $\mu$ is near the minimum value of the second
subband, as shown in Fig. 2(a) ($|E|=2.16$ eV).

\begin{figure}[h]
\centering
\includegraphics[angle=0,scale=0.3]{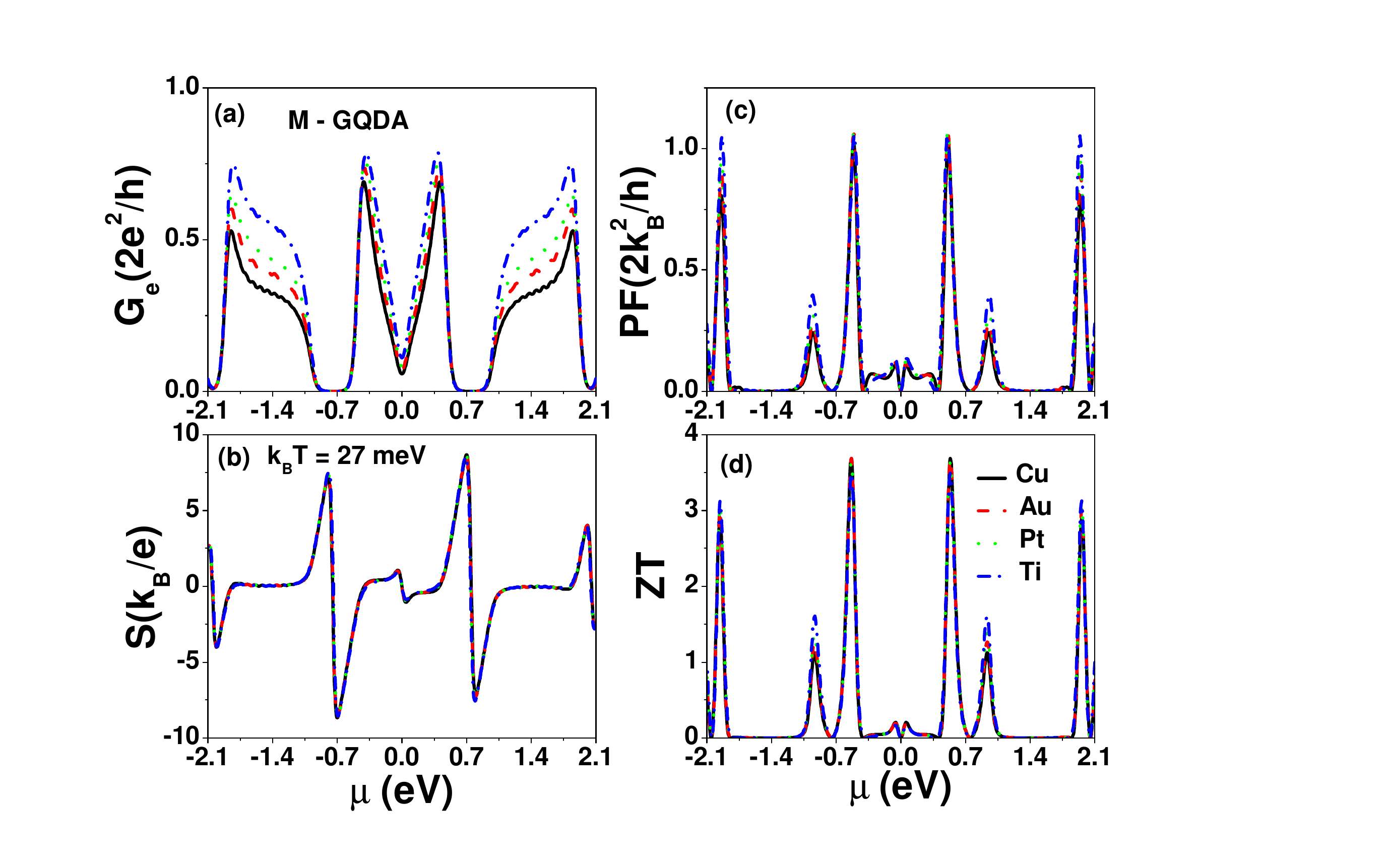}
\caption{(a) Electrical conductance $G_e$,(b) Seebeck coefficient
$S$, (c) power factor ($PF = S^2 G_e$), and (d) figure of merit
$ZT$ as functions of $\mu$ for various metal electrodes at $k_B
T=27~meV$.}
\end{figure}

\begin{figure}[h]
\centering
\includegraphics[angle=0,scale=0.3]{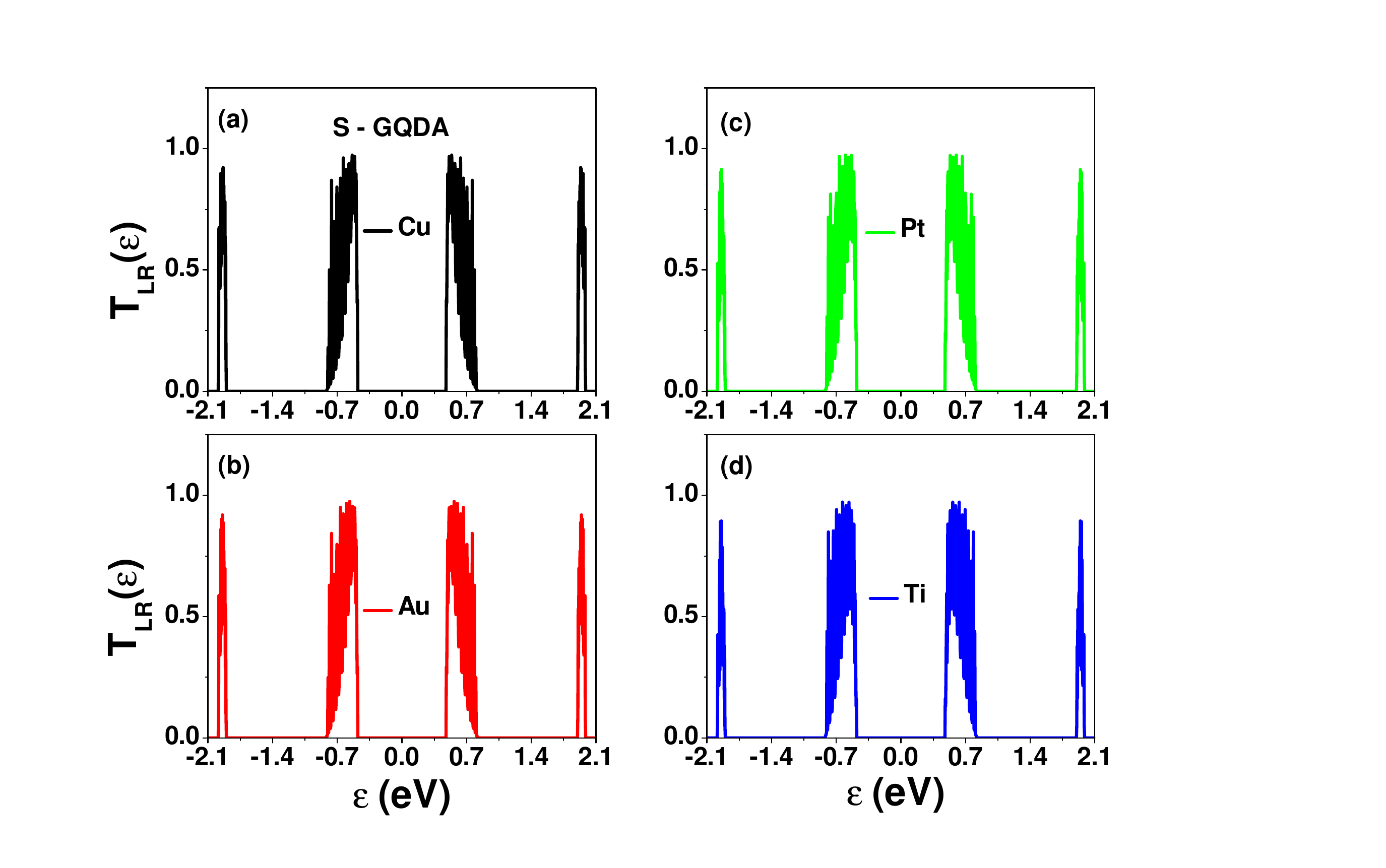}
\caption{Transmission coefficient ${\cal T}_{LR}(\varepsilon)$ of
a finite S-GQDA as functions of $\varepsilon$ for different
line-contacted metal electrodes at $N_a=8$ ( $L_a=0.71$nm ) and
$N_z=127$ ($L_z=15.5$nm ). Diagrams (a), (b), (c), and (d)
correspond to Cu, Au, Pt, and Ti, respectively. }
\end{figure}

In Fig. 6(b), the gaps between the minibands substantially enhance
the maximum $S$ values, suggesting that the thermoelectric
properties of S-GQDAs with a large central band gap should be
investigated. The calculated transmission coefficients of S-GQDAs
with various line-contacted metal electrodes are presented in Fig.
7. The miniband widths of S-GQDAs are much smaller than those of
M-GQDAs, resulting in a greatly reduced area of their ${\cal
T}_{LR}(\varepsilon)$ curves. Fig. 8 presents the calculated
$G_e$, $S$, power factor $PF$, and $ZT$ as functions of $\mu$ at
room temperature. Due to electron-hole symmetry, we considered the
range of chemical potential $\mu \ge 0$. The $G_e$ spectra of the
first miniband remains triangular as in Fig. 6(a), but the steep
change with respect to $\mu$ is toward the central gap. Because of
large gaps, the maximum Seebeck coefficients in Fig. 8(b) are
almost twice as large as those in  Fig. 6(b). The maximum power
factor $PF_{max}(\mu=0.45eV)=0.9682$ and figure of merit
$ZT_{max}(\mu=0.427eV)=3.442$ occur at the chemical potentials
near $\varepsilon_{e,1}=0.47905eV$ for Cu electrodes. The
$\kappa_{ph}$ value used in the analysis is the same as that in
Fig. 6(d). A comparison of the maximum $ZT$ in Fig. 8(d) and 6(d)
reveals that it did not increase despite large increase in the
maximum value of $S$. For $\kappa_{ph} \gg \kappa_e$, the increase
of $G_e$ also has a substantial effect. If the miniband width was
small,  $G_e$ was suppressed in the gap region. In Fig. 8(c) and
8(d), $PF_{max,2}$ and $ZT_{max,2}$ are also smaller than
$PF_{max,1}$ and $ZT_{max,1}$, respectively. These results
indicate that increasing  electrical power output requires not
only large gaps but also relatively wide miniband widths.

\begin{figure}[h]
\centering
\includegraphics[angle=0,scale=0.3]{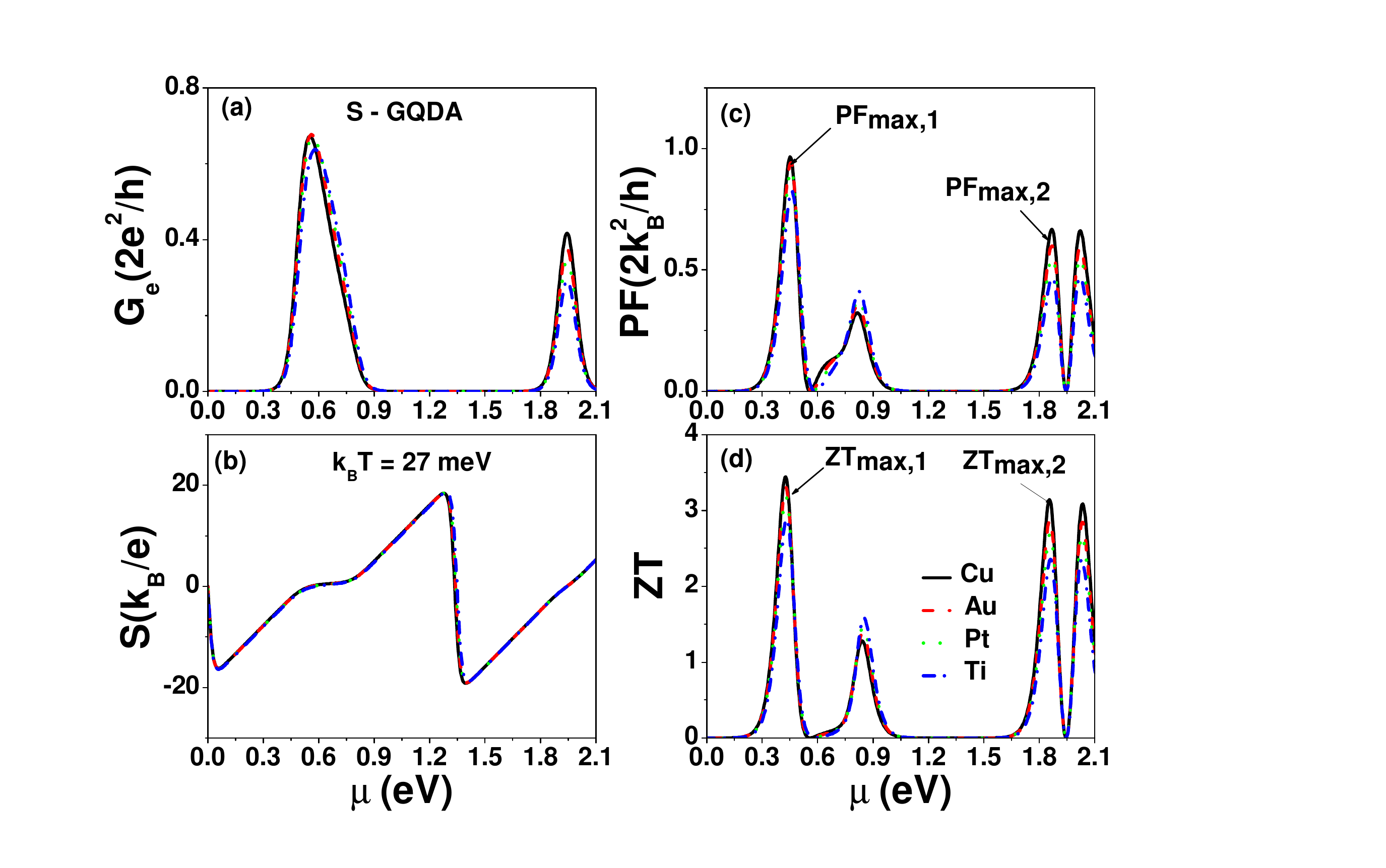}
\caption{(a) Electrical conductance $G_e$, (b) Seebeck coefficient
$S$,(c) power factor $PF=S^2G_e$ , and (d) figure of merit ($ZT$)
of S-GQDSLs as functions of $\mu$ for various metal electrodes at
$k_BT=27meV$.}
\end{figure}

\subsubsection{Zigzag edge atoms coupled to electrodes}
Our previous study demonstrated that the electron transport of
finite ZGNRs is significantly affected by the armchair or zigzag
edge sides coupled to electrodes [\onlinecite{KuoDMT}]. According
to [\onlinecite{LeeG}], most edge-oxidized modifications
dramatically change the electron band structures of ZGNRs.
Therefore, we do not consider edge-oxidized GNRs in the following
discussion. The contact resistance of Cr/X/graphene was calculated
in [\onlinecite{GaoQ}], in which  the zigzag edge atoms of
graphene were coupled to Cr by various atomic bridges (H, F, O, or
S). Due to the large binding distance, a high contact resistance
was observed in the Cr/H/graphene structure. We adopted
$\Gamma_t=9$ meV and $\Gamma_t=45$ meV for Cr/H/GQDAs and
Cr/F/GQDAs, respectively, to calculate $G_e$ as functions of $\mu$
at zero temperature (Fig. 9) for  M-GQDAs and S-GQDAs with $N_a=8$
and $N_z=127$. In Fig. 9, the spectra of $G_e$ are larger than
quantum conductance $G_0=\frac{2e^2}{h}$, demonstrating that GQDAs
behave as parallel QDs when the zigzag edge atoms of M-GQDAs and
S-GQDAs are coupled to the metal electrodes. In the parallel QD
configuration, the spectra resulting from electronic states near
the zero-energy modes of M-GQDAs and the high-energy modes of the
first minibands of S-GQDAs substantially differ from the spectra
in Fig. 5 and Fig. 7.

\begin{figure}[h]
\centering
\includegraphics[angle=0,scale=0.3]{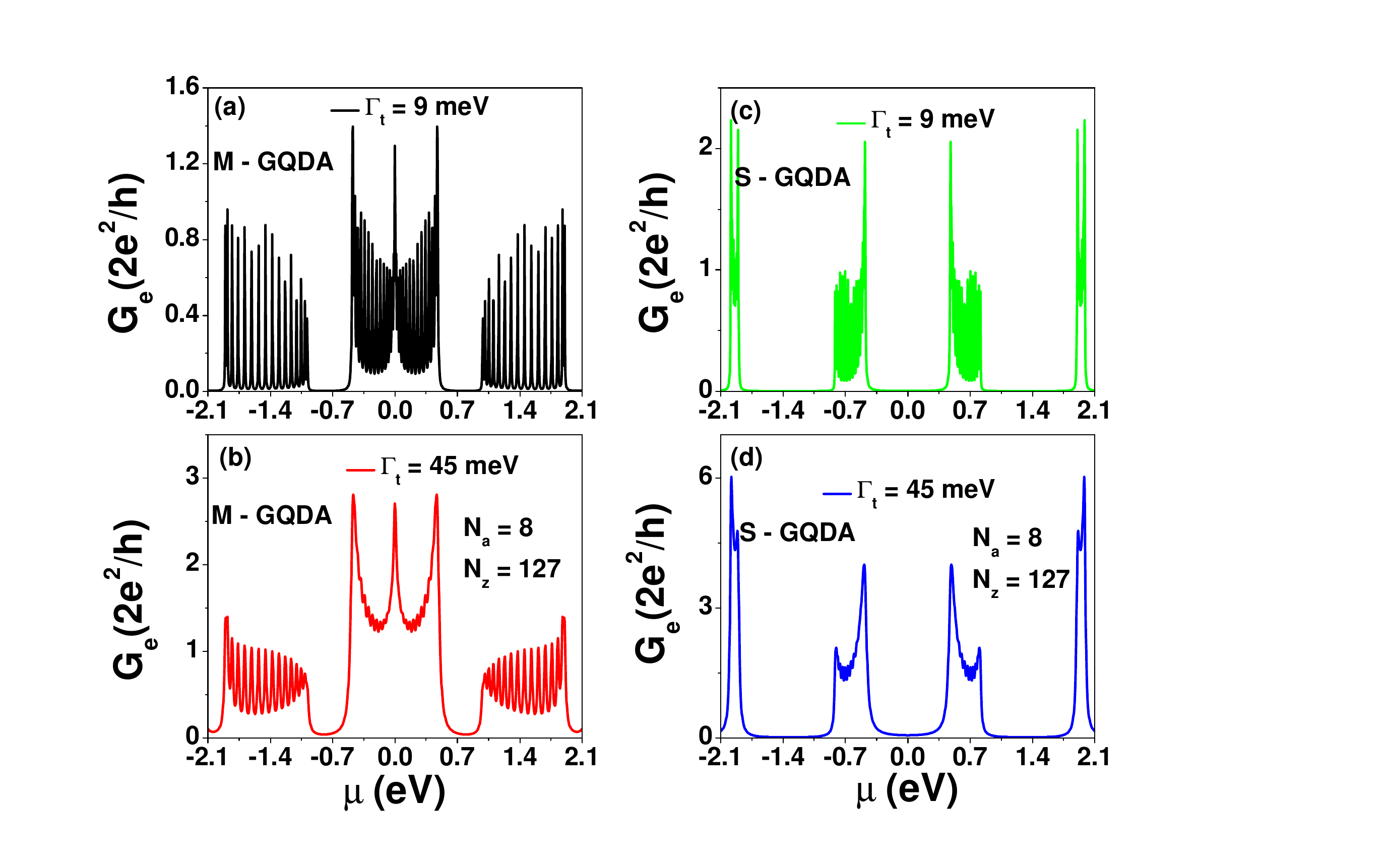}
\caption{Electrical conductance as functions of $\mu$ for
$\Gamma_t=9$ meV and $\Gamma_t=45$ meV for (a,b)  M-GQDAs and
(c,d) S-GQDAs. Their sizes are $N_a=8$ ($L_a=0.71$ nm) and
$N_z=127$ ($L_z=15.5$ nm).}
\end{figure}

Fig. 10 presents  the calculated $S$ and $PF$ of M-GQDAs and
S-GQDAs at room temperature ($k_BT=27$ meV). The maximum $S$ is
severely degraded for a parallel QD configuration with a short
channel length of $L_a=0.71$ nm compared with the maximum $S$
values of Fig. 6 and Fig. 8. This is attributable to the highly
enhanced $G_e$ in the gap region, which is similar to the
metal-induced gap states in metal-semiconductor Schottky junctions
[\onlinecite{Roksana}]. When the zigzag edge atoms are coupled to
the metal electrodes, the miniband edge states are readily
extended into the gap regions if the channel length is short.
Although the maximum $PF$ in Fig. 10 is much larger than
$PF_{QB}=1.2659(\frac{2k^2_B}{h})$, their $ZT$ values are less
than one due to remarkable increases in  electron thermal
conductance ($\kappa_e$) and phonon thermal conductance
($\kappa_{ph}$).

\begin{figure}[h]
\centering
\includegraphics[angle=0,scale=0.3]{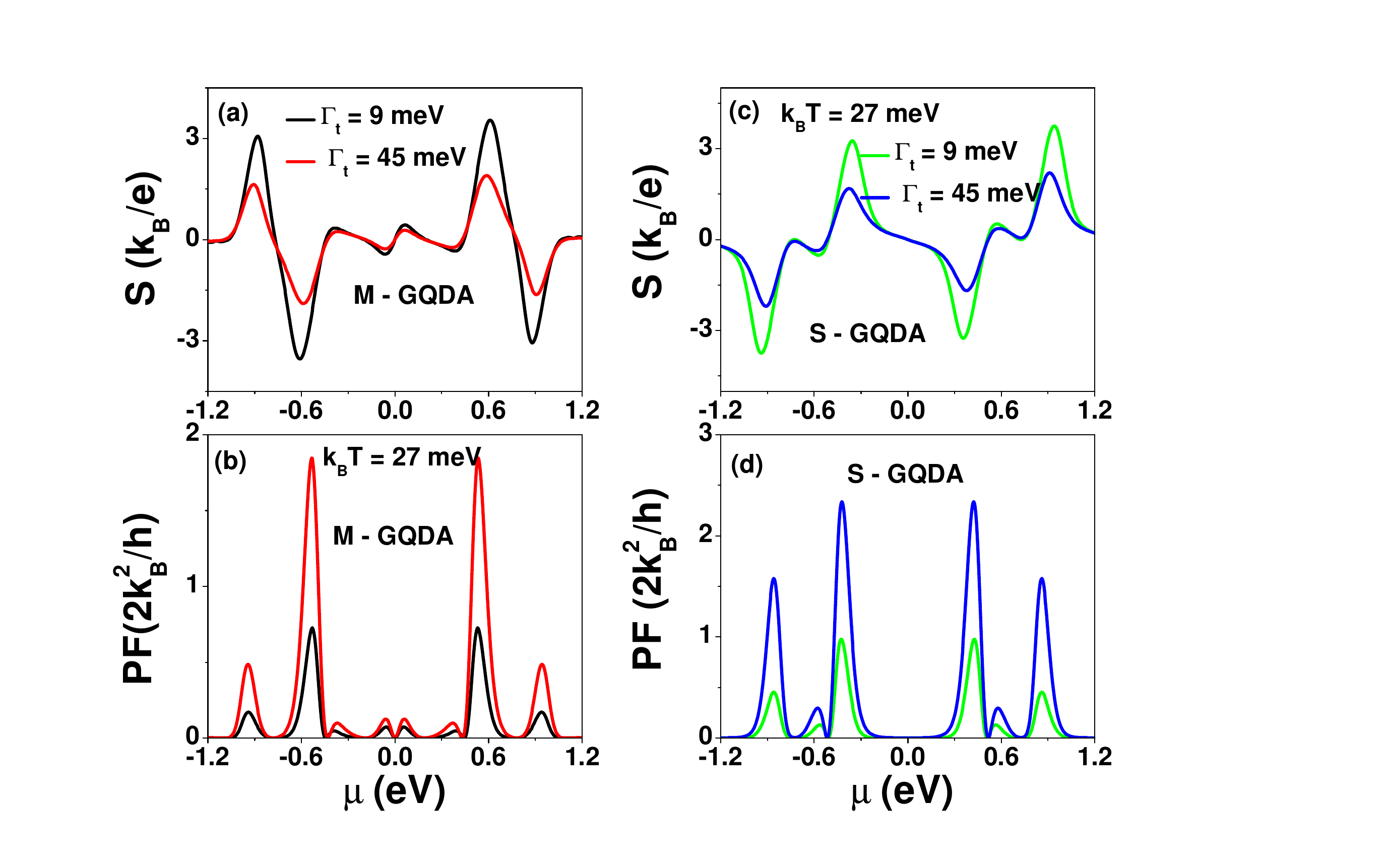}
\caption{(a) Seebeck coefficient and (b) power factor of M-GQDAs
and (c,d) S-GQDAs as functions of $\mu$ for two tunneling rates at
$k_BT=27$meV. Other physical parameters are the same as those in
Fig. 9.}
\end{figure}

\subsection{Surface-contacted  metal electrodes}
Few experimental studies have reported the transport and
thermoelectric properties of GNRs due to the difficulty of
applying the line-contacted technique [\onlinecite{CustiT}].
However, surface-contacted metal electrodes are straightforward to
lay out in an experiment. Therefore, experimental methods are
preferable for revealing the thermoelectric properties of GNRs
with surface-contacted electrodes. Many theoretical studies
 have demonstrated that the
surface contact resistance is one to three orders of magnitude
greater than the line contact resistance [\onlinecite{Matsuda},
\onlinecite{ChuT}-- \onlinecite{Hancheng}], indicating that
$\Gamma_t$ may be extremely small. The calculated transmission
coefficient of M-GQDA as functions of $\mu$ for various $\Gamma_t$
values is presented in Fig. 11;  the two outer GQDs are coupled to
metal electrodes. The structure has 28 carbon atoms, including 6
carbon atoms each located at zigzag edge sites under the
surface-contacted left (right) metal electrode. Due to the small
$\Gamma_t$ values, the peaks of the  ${\cal T}_{LR}(\varepsilon)$
spectra are very narrow. These molecule-like spectra indicate
degraded electrical conductance at room temperature.

\begin{figure}[h]
\centering
\includegraphics[angle=0,scale=0.3]{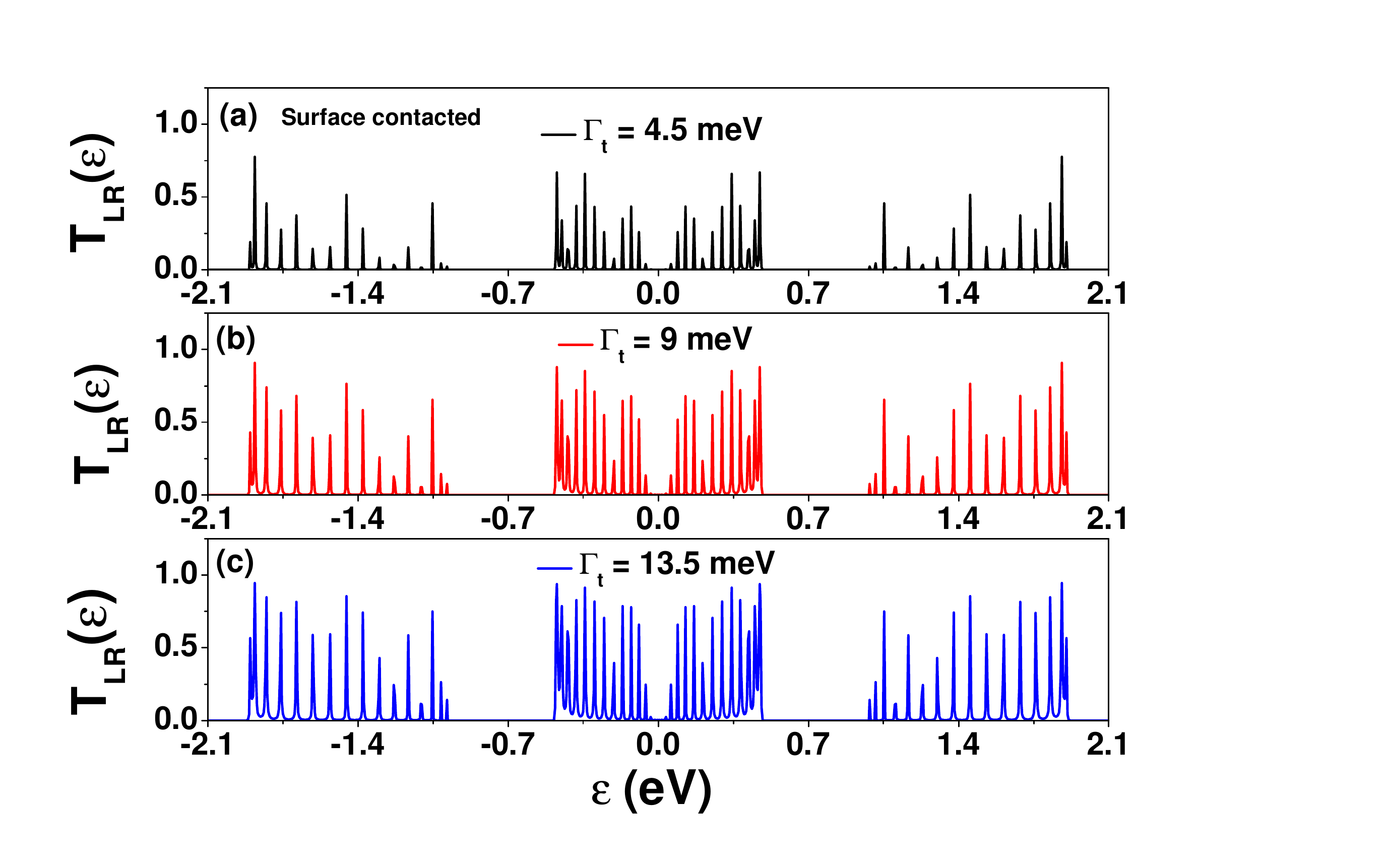}
\caption{Transmission coefficient of an  M-GQDA with
surface-contacted metal electrodes and  $N_a=8$ and $N_z=127$ as
functions of $\varepsilon$ for various $\Gamma_t$ values. (a),
(b),  and (c) correspond to $\Gamma_t = 4.5$, $9$, and $13.5 $
meV, respectively.}
\end{figure}

Fig. 12 presents the calculated $G_e$, $\kappa_e$, $PF$, and $ZT$
as functions of $\mu$ at different temperatures with
surface-contacted metal electrodes. The maximum electrical
conductance $G_e=0.25G_0$ at room temperature is much smaller than
that in Fig.6(a). Unlike the behavior of $G_e$ with respect to
temperature, electron thermal conductance $\kappa_e$ is enhanced
as the temperature increases. Here, $\kappa_e$ is in units of
$\kappa_0=0.62nW/K$. The increase in the maximum $PF$ with
decreasing temperature is attributable to the enhancement of $S$,
which is a highly nonlinear function of temperature. In the
calculation of $ZT$, $\kappa_{ph}$ is the same as that in Fig. 5.
At room temperature, the maximum value of $ZT(\mu=0.535eV)=1.9927$
is less than three. However, the maximum $PF$ and $ZT$ values
increase with decreasing temperature. If the spectra of $G_e$ have
molecule-like characteristics, the thermoelectric performance of
GQDAs remains acceptable in the low-temperature region
[\onlinecite{Kuo1}].

\begin{figure}[h]
\centering
\includegraphics[angle=0,scale=0.3]{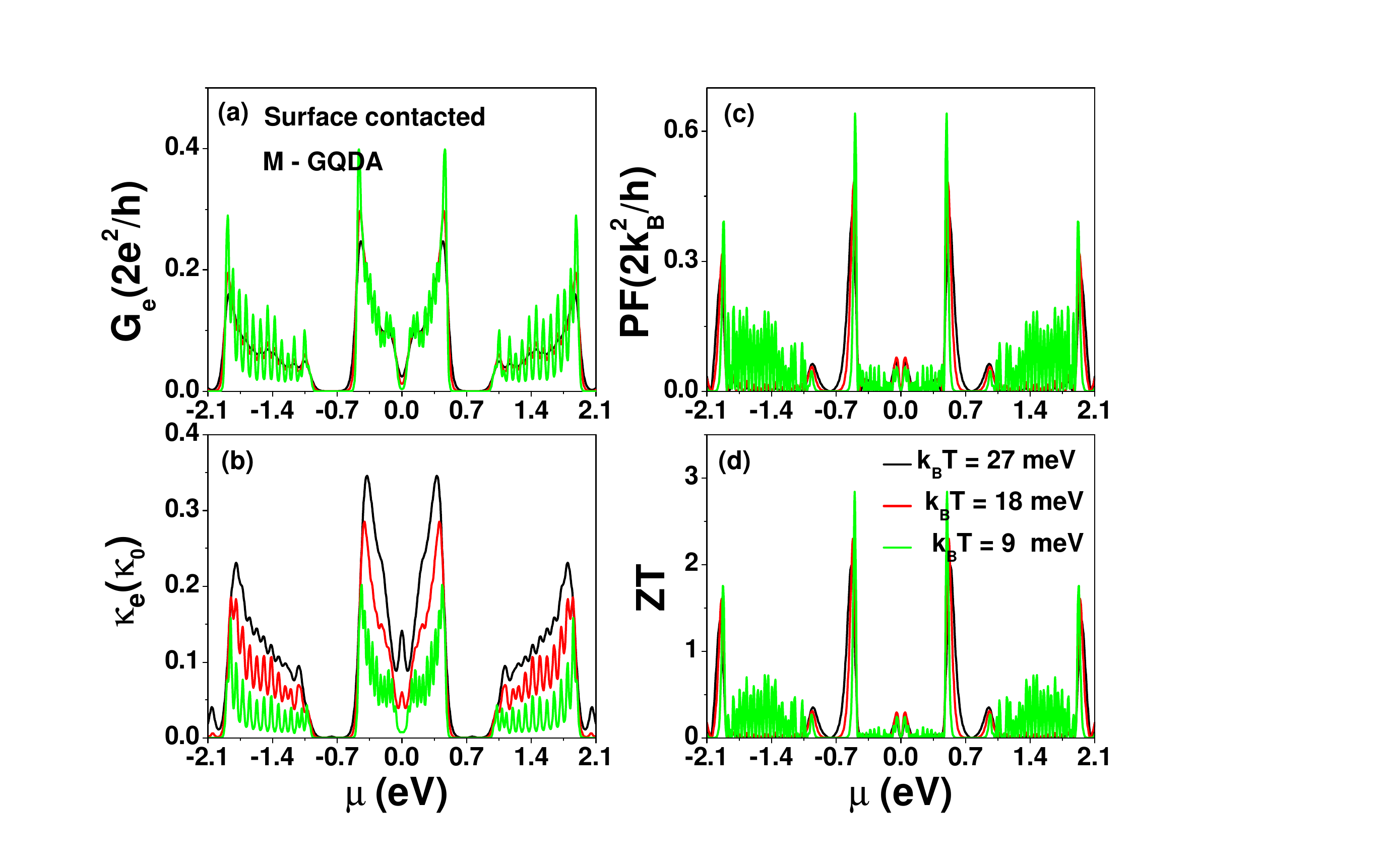}
\caption{(a) Electrical conductance, (b) electron thermal
conductance, (c) power factor, and (d) figure of merit as
functions of $\mu$ for various temperatures for M-GQDA with
surface-contacted  metal electrodes. $\Gamma_t=14.5$ meV. Other
physical parameters are the same as in Fig. 11.}
\end{figure}

The molecule-like or band-like spectra of GQDAs are determined by
not only the lengths of the GQDAs but also the contact geometry.
In the line-contacted metal electrodes, the spectra of finite
GQDAs have band-like characteristics due to the large $\Gamma_t$
values. By contrast, the spectra of finite GQDAs with
surface-contacted metal electrodes have molecule-like
characteristics. We did not investigate the effect of
electron--electron  Coulomb interactions in this study. Although
electron Coulomb interactions are strong in small GQDAs
($L_a=0.71$ nm and $L_z=15.5$ nm), their effects on the
thermoelectric quantities can be neglected if electron transport
is dominated by the thermionic-assisted tunneling procedure
(TATP), which occurs at a distance of $\mu$ away from the
minibands. The thermionic electron population in the minibands is
very dilute if  $\mu$ is located  inside the gap. Hence, electron
correlation functions arising from electron Coulomb interactions
are small [\onlinecite{Kuo1}--\onlinecite{Kuo3}]. The calculated
maximum $PF$ and $ZT$ are therefore valid in the TATP.

\section{Conclusion}
Although ZGNRs have metallic phases, the minibands and gaps are
formed in ZGNRs with periodic vacancies, which behave as GQDAs.
Such GQDAs can host metallic and semiconductor phases. The edge
states of the first conductance and valence minibands, such as
$\varepsilon_{e(h),1}$, are localized edge states along the
armchair edge direction, whereas they have very long coherent
lengths along the zigzag edge direction. The designed GQDAs have
sufficiently large gaps and miniband widths to be applied in
thermoelectric devices operating at room temperature. With
line-contacted  metal electrodes, the maximum power factor (figure
of merit) of the M-GQDAs at room temperature is larger than that
of the S-GQDAs due to the  wider miniband widths of the M-GQDAs.
The GQDAs behave as  serially coupled QDs (SCQDs) if the armchair
edge atoms are coupled to the metal electrodes. By contrast, the
 GQDAs behave as parallel-coupled
QDs if zigzag edge atoms are coupled to metal electrodes. For
SCQDs with line-contacted  electrodes, the maximum power factor
values of GQDAs with different metal electrodes (Cu, Au, Pt, Pd,
or Ti) at room temperature can reach $79\%$ of the theoretical
limit of 1D systems, and their figures of merit are greater than
three. If surface-contacted metal electrodes are used, the spectra
of electrical conductance of the finite M-GQDAs have molecule-like
characteristics because of high contact-surface resistance. The
thermoelectric power factor and figures of merit of the M-GQDAs at
room temperature are smaller than those of the line-contacted
M-GQDAs; however, their ZT values can reach three at low
temperatures.


{}

{\bf Acknowledgments}\\

We thank Yia-Chung Chang for their valuable intellectual input.

This work was supported by the Ministry of Science and Technology
(MOST), Taiwan under Contract No. 110-2119-M-008-006-MBK.

\mbox{}\\
E-mail address: mtkuo@ee.ncu.edu.tw\\

 \numberwithin{figure}{section} \numberwithin{equation}{section}

\setcounter{section}{0}
\setcounter{equation}{0} 

\mbox{}\\






\newpage

\begin{thebibliography}{100}



\bibitem[1]{HaugH}  Haug H, and  Jauho A P, Quantum Kinetics in Transport and
Optics of Semiconductors (Springer, Heidelberg, 1996)

\bibitem[2]{GuoLJ}  Guo L J,  Leobandung E and Chou S Y, A silicon single-electron transistor memory operating at room temperature,
1997 Science \textbf{275} 649

\bibitem[3]{Postma} Postma H W C, Teepen T, Yao Z,
Grifoni M and Dekker C, Carbon nanotube single-electron
transistors at room temperature, 2001 Science \textbf{293} 76


\bibitem[4]{Michler}  Michler P, Imamoglu A,  Mason M D, Carson  P J,
Strouse G F and Buratto S K, Quantum correlation among photons
from a single quantum dot at room temperature, 2000 Nature
\textbf{406} 968

\bibitem[5]{Santori}  Santori C,  Fattal D, Vuckovic J, Solomon G S and
Yamamoto Y, Indistinguishable photons from a single-photon device,
2002 Nature \textbf{419} 549

\bibitem[6]{ChangWH}  Chang W H, Chen W Y,  Chang H S, Hsieh T P,
and Hsu T M, Efficient single-photon sources based on low-density
quantum dots in photonic-crystal nanocavities, 2006 Phys. Rev.
Lett. \textbf{96} 117401

\bibitem[7]{Gustavsson}  Gustavsson S,  Studer M,  Leturcq R,  Ihn T,  Ensslin K, Driscoll D
C and  Gossard A C, Frequency-selective single-photon detection
using a double quantum dot, 2007 Phys. Rev. Lett. \textbf{99}
206804

\bibitem[8]{Josefsson} Josefsson  M,  Svilans A,  Burke A M,  Hoffmann E A,
Fahlvik  S, Thelander C, Leijnse M and Linke H, A quantum-dot heat
engine operating close to the thermodynamic efficiency limits,
2018 Nature Nanotechnology \textbf{13} 920

\bibitem[9]{Kagan}  Kagan C R and Murry C B, Charge transport in strongly coupled quantum dot solids,
2015 Nature Nanotechnology \textbf{10} 1013

\bibitem[10]{Harman}  Harman T C,  Taylor P J,  Walsh M P and
LaForge B E, Quantum dot superlattice thermoelectric materials and
devices, 2002 Science \textbf{297} 2229

\bibitem[11]{Talgorn}  Talgorn E,  Gao Y,  Aerts M,  Kunneman L T, Schins J M,
Savenije T J, Marijn A. van Huis, Herre S. J. van der Zant,
Houtepen Arjan J and Siebbeles Laurens D A, Unity quantum yield of
photogenerated charges and band-like transport in quantum-dot
solids, 2011 Nature Nanotechnology \textbf{6}733


\bibitem[12]{Lawrie} Lawrie W I L, Eenink H G J,  Hendrickx N W,  Boter J M, Petit L,
Amitonov S V, Lodari  M,  Wuetz Paquelet B,  Volk C. and Philips S
G J et al, Quantum dot arrays in silicon and germanium, 2020 Appl.
Phys. Lett. \textbf{116} 080501

\bibitem[13]{Novoselovks}  Novoselov K S, Geim A K, Morozov S V, Jiang D, Zhang Y,
Dubonos S V, Grigorieva  I V and  Firsov A A, Electric field
effect in atomically thin carbon films, 2004 Science \textbf{306}
666

\bibitem[14]{Geim} Geim A K and  Grigorieva I V, Van der Waals heterostructures, 2013 Nature \textbf{499} 419

\bibitem[15]{NovoselovKS} Novoselov K S,  Mishchenko A, Carvalho A and
Neto A H C, 2D materials and van der Waals heterostructures, 2016
Science \textbf{353} aac9439

\bibitem[16]{Desai}  Desai S B, Seol G, Kang J S, Fang H, Battaglia C,
 Kapadia R, Ager J W,  Guo J and Javey A, Strain-Induced
Indirect to Direct Bandgap Transition in Multi layer WSe2, 2014
Nano Lett. \textbf{14} 4592

\bibitem[17]{Cai}  Cai J,  Ruffieux P,  Jaafar R,  Bieri M, Braun T,  Blankenburg S,  Muoth M,
Seitsonen A P,  Saleh M,  Feng X,  Mullen K and  Fasel Roman,
Atomically precise bottom-up fabrication of graphene nanoribbons,
2010 Nature \textbf{466} 470


\bibitem[18]{LiuJ}  Liu J Z, Li B W,
Tan Y Z,  Giannakopoulos A,  Sanchez-Sanchez C,  Beljonne D,
Ruffieux P, Fasel R, Feng X L and Mullen K, Toward Cove-Edged Low
Band Gap Graphene Nanoribbons, 2015 J. Am. Chem. Soc. \textbf{137}
6097

\bibitem[19]{ChenYC}  Chen Y C,  Cao T,  Chen C,  Pedramraz Z,
Haberer D, Oteyza  D G de, F.  Fischer R,  Louie S G and Crommie M
F, Molecular bandgap engineering of bottom-up synthesized graphene
nanoribbon heterojunctions, 2015 Nature Nanotechnology \textbf{10}
156

\bibitem[20]{RuffieuxP}  Ruffieux P, Wang S,  Yang B,  Sachez-Sachez C,  Liu J,  Dienel T,  Talirz L,
Shinde P,  Pignedoli C A, Passerone D, Dumslaff T,  Feng X, Mullen
K and Fasel R, On-surface synthesis of graphene nanoribbons with
zigzag edge topology, 2016 Nature \textbf{531} 489

\bibitem[21]{Groning} Groning  O,  Wang S,  Yao X, Pignedoli C
A,  Barin G B,  Daniels C,  Cupo A,  Meunier V,  Feng X, Narita A,
Muellen K, Ruffieux P and  Fasel R, Engineering of robust
topological quantum phases in graphene nanoribbons, 2018 Nature
\textbf{560} 209

\bibitem[22]{Rizzo} Rizzo  D J, Veber  G, Cao T,  Bronner C,  Chen T,
Zhao F, Rodriguez H,  Louie S G,  Crommie M F and  Fischer F R,
Topological band engineering of graphene nanoribbons, 2018 Nature
\textbf{560} 204

\bibitem[23]{Yan} Yan L H and  Liljeroth P, Engineered electronic states in atomically precise artificial lattices and graphene
nanoribbons, 2019 ADVANCES IN PHYSICS: X \textbf{4} 1651672

\bibitem[24]{DRizzo} Rizzo D J, Veber G,  Jiang J W,  McCurdy R, Bronner
T, Cao T, Chen T,  Louie Steven G, Fischer F R and Crommie M F,
Inducing metallicity in graphene nanoribbons via zero-mode
superlattices, 2020 Science \textbf{369} 1597

\bibitem[25]{SunQ}  Sun Q,  Yan Y,  Yao X L, Mullen K,  Narita A, Fasel R and  Ruffieux P, Evolution of the Topological Energy Band in Graphene Nanoribbons
, 2021 J. Phys. Chem. Lett. \textbf{12} 8679

\bibitem[26]{DJRizzo}  Rizzo D J, Jiang J W, Joshi D,  Veber G, Bronner C,  Durr R A,
 Jacobse P H,  Cao T,  Kalayjian A,  Rodriguez H,  Butler P,
Chen T,  Louie Steven G,  Fischer F R and  Crommie M F, Rationally
Designed Topological Quantum Dots in Bottom-Up Graphene
Nanoribbons, 2021 ACS Nano \textbf{15} 20633


\bibitem[27]{WangX}  Wang X, Ma J,  Zheng W H,  Osella S,
 Arisnabarreta N,  Droste J, Serra J, Ivasenko O,  Lucotti A,
Beljonne D, Bonn M,  Liu X Y,  Hansen M R,  Tommasini M, Feyter S
De,  Liu J Z, Wang H I and  Feng  X L, Cove-Edged Graphene
Nanoribbons with Incorporation of Periodic Zigzag-Edge Segments,
2022 J. Am. Chem. Soc. \textbf{144} 228

\bibitem[28]{Almeida}  Almeida P A and  Martins G B, Thermoelectric transport properties of armchair graphene nanoribbon heterostructures,
2022 J. Phys: Condens. Matter \textbf{34} 335302

\bibitem[29]{Topsakal} Topsakal M,  Sevincli H and Ciraci S, Spin confinement in the superlattices of graphene
ribbons, 2008 Appl. Phys. Lett. \textbf{92} 173118




\bibitem[30]{Cuniberti} Sevincli H and  Cuniberti G, Enhanced thermoelectric figure of merit in edge-disordered zigzag graphene nanoribbons, 2010 Phys. Rev.
B \textbf{81} 113401

\bibitem[31]{Darancet}  Darancet P,  Olevano V and  Mayou D, Coherent Electronic Transport through Graphene Constrictions: Subwavelength Regime and Optical Analogy,
2009 Phys. Rev. Lett. \textbf{102} 136803

\bibitem[32]{GLiang}  Liang G C,  Neophytou N,  Lundstrom M S
and  Nikonov D E, Contact effects in graphene nanoribbon
transistors, 2008 Nano. Lett. \textbf{8} 1819


\bibitem[33]{Matsuda} Matsuda Y,  Deng W Q and  Goddard
III W A, Contact Resistance for "End-Contacted" Metal-Graphene and
Metal-Nanotube Interfaces from Quantum Mechanics, 2010 J. Phys.
Chem. C. \textbf{114} 17845


\bibitem[34]{Shen} Shen P C, Su C,  Lin Y X,
Chou A S, Cheng C C,  Park J H,  Chiu M H, Lu A Y,  Tang H L,
Tavakoli M M, Pitner G, Ji X,  Cai Z Y, Mao N N, Wang J T, Tung J
V,  Li C,  Bokor J, Zettl A,  Wu C I, Palacios  T, Li L J and Kong
J, Ultralow contact resistance between semimetal and monolayer
semiconductors, 2021 Nature \textbf{593} 212

\bibitem[35]{ChenRS}  Chen R S, Ding  G L,  Zhou Y and Han S T, Fermi-level depinning of 2D transition metal dichalcogenide transistors,
2021 J. Mater. Chem. C, \textbf{9} 11407

\bibitem[36]{Wakabayashi2}  Wakabayashi K, Sasaki K,
Nakanishi T and Enoki T, Electronic states of graphene nanoribbons
and analytical solutions, 2010 Sci. Technol. Adv. Mater.
\textbf{11} 054504

\bibitem[37]{KuoDM}  Kuo D M T, Thermoelectric and electron heat rectification properties of quantum dot superlattice nanowire arrays,
2020 AIP Advances \textbf{10} 045222



\bibitem[38]{KuoDMT}  Kuo D M T, Effects of zigzag edge states on the thermoelectric properties of finite graphene nanoribbons,
2022 Jpn. J. Appl. Phys. \textbf{61} 075001

\bibitem[39]{KuoD} Kuo D M T and  Chang Y C, Contact Effects on Thermoelectric Properties of Textured Graphene Nanoribbons,
2022 Nanomaterials \textbf{12} 3357

\bibitem[40]{XuY} Xu Y, Li Z Y, and Duan W H, Thermal and
thermoelectric properties of graphene, 2014 Small \textbf{10} 2182

\bibitem[41]{Nakada}  Nakada K, Fujita M, Dresselhaus G and  Dresselhaus M S,
Edge state in graphene ribbons: Nanometer size effect and edge
shape dependence, 1996 Phys. Rev. B \textbf{54} 17954

\bibitem[42]{Wakabayashi}  Wakabayashi K,  Fujita M,  Ajiki H and  Sigrist M, Electronic and magnetic properties of nanographite ribbons,
1999 Phys. Rev. B \textbf{59} 8271


\bibitem[43]{Gunst} Gunst T, Markussen T,  Jauho A P, and Brandbyge M,
Thermoelectric properties of finite graphene antidot lattices,
2011 Phys. Rev. B \textbf{84} 155449

\bibitem[44]{ChangPH}  Chang P H,  Bahramy M S,  Nagaosa N and
Nikolic B K, Giant Thermoelectric Effect in Graphene-Based
Topological Insulators with Heavy Adatoms and Nanopores, 2014 Nano
Lett. \textbf{14} 3779

\bibitem[45]{ZhangYT} Zhang Y T, Li Q M, Li Y C, Zhang Y Y
and  Zhai F, Band structures and transport properties of zigzag
graphene nanoribbons with antidot array, 2010 J. Phys: Condens.
Matter \textbf{22} 315304




\bibitem[46]{ChuT} Chu T and Chen Z, Understanding the Electrical Impact of
Edge Contacts in Few-Layer Graphene, 2014 ACS Nano \textbf{8} 3584


\bibitem[47]{GongC} Gong C,  McDonnell S,  Qin X Y,  Azcatl A,  Dong H,  Chabal Y J,  Cho K, and  Wallace R M,
Realistic Metal-Graphene contact structures, 2014 ACS Nano
\textbf{8} 642

\bibitem[48]{SongSM} Song S M, Kim T Y,  Sul O J, Shin W C and Cho B J,
Improvement of graphene-metal contact resistance by introducing
edge contacts at graphene under metal, 2014 Appl. Phys. Lett.
\textbf{104} 183506

\bibitem[49]{GaoQ} Gao Q and  Guo J, Role of chemical termination in edge contact
to graphene, 2014 APL Mater. \textbf{2} 056105

\bibitem[50]{MatsudaY1} Matsuda Y,  Deng W Q, and Goddard III W A, Contact resistance properties
between nanotubes and various metals from quantum mechanics, 2007
J. Phys. Chem. C \textbf{111} 11113


\bibitem[51]{Massen}  Massen J,  Ji W, and  Guo H, First principles study of
electronic transport through a Cu(111)/graphene junction. 2010
Appl. Phys. Lett. \textbf{97} 142105

\bibitem[52]{Barraza}  Barraza-Lopez S, Vanevic M, Kindermann M and Chou M Y,
Effects of Metallic Contacts on Electron Transport through
Graphene, 2010 Phys. Rev. Lett. \textbf{104} 076807

\bibitem[53]{ShenC}  Shen C, Liu J, Jiao N, Zhang C X,  Xiao H, Wang R Z and Sun L Z, Transport properties of graphene/metal planar
junction, 2014 Phys. Lett. A \textbf{378} 1321


\bibitem[54]{RanQ}  Ran Q,  Gao M,  Guan X,  Wang Y and  Yu Z, First principles
investigation on bonding formation and electronic structure of
metal graphene contacts, 2009 Appl. Phys. Lett. \textbf{94} 103511


\bibitem[55]{StokbroK} Stokbro K, Engelund M and  Blom A, Atomic-scale model for
the contact resistance of the nickel-graphene interface, 2012
Phys. Rev. B\textbf{ 85} 165442

\bibitem[56]{MaB} Ma  B,  Cheng G, Wen Y W, Chen R,  Cho K G and  Shan B, Modulation of contact resistance between metal
and graphene by controlling the graphene edge, contact area, and
points defects: An ab initio study, 2014 J. Appl. Phys.
\textbf{115} 183708

\bibitem[57]{Hancheng} Qin H C, Lu W C and Bernholca J, Ab initio simulations of metal contacts for
graphene-based devices, 2022 J. Appl. Phys. \textbf{131} 214301

\bibitem[58]{YangL} Yang L, Park C H, Son Y W, Cohen Marvin L, and
Louie Steven G, Quasiparticle Energies and Band Gaps in Graphene
Nanoribbons, 2007 Phys. Rev. Lett. \textbf{99} 186801


\bibitem[59]{LeeG}  Lee G and Cho  K, Electronic structures of zigzag graphene nanoribbons with edge hydrogenation and
oxidation, 2009 Phys. Rev. B \textbf{79} 165440


\bibitem[60]{Whitney} R. S. Whitney, Most Efficient Quantum Thermoelectric at Finite Power Output,
2014 Phys. Rev. Lett.\textbf{112} 130601


\bibitem[61]{Zhengh} Zheng H, Liu H J, Tan X J,  Lv H Y, Pan L, Shi J and Tang X
F, Enhanced thermoelectric performance of graphene nanoribbons,
2012 Appl. Phys. Lett. \textbf{100} 093104

\bibitem[62]{Xu} Xu  Y, Gan  Z and Zhang S C,
Enhanced Thermoelectric Performance and Anomalous Seebeck Effects
in Topological Insulators, 2014 Phys. Rev. Lett. \textbf{112}
226801


\bibitem[63]{Roksana} Golizadeh-Mojarad R and Datta S, Effect of contact
induced states on minimum conductivity in graphene, 2009 Phys.
Rev. B \textbf{79} 085410


\bibitem[64]{CustiT}  Cusati T, Fiori G, Gahoi A,
 Passi V, Lemme M C, Fortunelli A, and Iannaccone G,
Electrical properties of graphene-metal contacts, 2017 Scientific
Reports \textbf{7} 5109

\bibitem[65]{Kuo1} Kuo D M T and  Chang Y C,
Thermoelectric and thermal rectification properties of quantum dot
junctions, 2010 Phys. Rev. B \textbf{81} 205321

\bibitem[66]{Kuo6} Kuo D M T, Shiau S Y and Chang Y C
, Theory of spin blockade, charge ratchet effect, and
thermoelectrical behavior in serially coupled quantum dot system,
2011 Phys. Rev. B \textbf{84} 245303

\bibitem[67]{Kuo3}  Kuo D M T, Chen C C and Chang Y C,
Large enhancement in thermoelectric efficiency of quantum dot
junctions due to increase of level degeneracy, 2017 Phys. Rev. B
\textbf{95} 075432



\end{thebibliography}
\end{document}